\documentclass[reprint,amsmath,amssymb,aps,prb]{revtex4-1}
\usepackage{mathptmx}

\usepackage{natbib}

\usepackage{amsmath}
\usepackage{amsthm}
\usepackage{bm}
\usepackage{bbm}

\usepackage{graphicx}

\usepackage[caption=false]{subfig}

\usepackage{algpseudocode}
\usepackage{algorithm}

\usepackage{booktabs}
\usepackage{multirow}

\usepackage[per=reciprocal,obeyfamily,obeybold,obeyitalic]{siunitx} 
\usepackage{ragged2e}

\usepackage{xspace}

\usepackage[colorlinks=false, pdfborder={0 0 0}]{hyperref}
\usepackage{cleveref}

\newcommand{\dd}{\mathrm{d}}
\newcommand{\half}{\frac{1}{2}}

\newcommand{\deriv}[2]{\frac{\dd #1}{\dd #2}}

\newcommand{\SIESTA}{\textsc{Siesta}\xspace}

\newcommand{\MD}{\textsc{md}\xspace}

\newcommand{\DFT}{\textsc{dft}\xspace}

\newcommand{\dirnfamily}[1]{\ensuremath{\langle \text{#1} \rangle}}

\usepackage{dcolumn}

\begin{document}

\title{Ab initio calculation of the shock Hugoniot of bulk silicon}%
\author{Oliver Strickson}
\email{ots22@cam.ac.uk}
\affiliation{Laboratory for Scientific Computing, Cavendish Laboratory, University of Cambridge, J. J. Thomson Avenue, Cambridge CB3 0HE, United Kingdom}
\author{Emilio Artacho}
\affiliation{Theory of Condensed Matter, Cavendish Laboratory, University of Cambridge, J. J. Thomson Avenue, Cambridge CB3 0HE, United Kingdom}
\affiliation{CIC Nanogune and DIPC, Tolosa Hiribidea 76, 20018 San Sebasti\'an, Spain}
\affiliation{Basque Foundation for Science Ikerbasque, Bilbao, Spain}
\date{July 2015}

\begin{abstract}
  We describe a simple annealing procedure to obtain the Hugoniot
  locus (states accessible by a shock wave) for a given material in a
  computationally efficient manner.  We apply this method to determine
  the Hugoniot locus in bulk silicon from ab~initio molecular dynamics
  with forces from density-functional theory, up to \SI{70}{GPa}.  The
  fact that shock waves can split into multiple waves due to phase
  transitions or yielding is taken into account here by specifying the
  strength of any preceding waves explicitly based on their yield
  strain. Points corresponding to uniaxial elastic compression along
  three crystal axes and a number of post-shock phases are given,
  including a plastically-yielded state, approximated by an isotropic
  stress configuration following an elastic wave of predetermined
  strength.  The results compare well to existing experimental data
  for shocked silicon.
\end{abstract}

\maketitle

\section{Introduction}

Shock waves are used extensively to study matter at conditions of
extreme pressure and temperature, and have been used to obtain some of
the highest laboratory-attained pressures.  They are useful for
equation of state determination and are important dynamic phenomena in
their own right, arising in aerodynamics,\cite{Dolling2001} reactive
flow\cite{Dlott2011} and high-speed impact.\cite{Duvall1977,Asay1993}

Simulations of shock waves have a long history.\cite{Holian2004}
Direct simulations using empirical potentials are now feasible on a
multi-billion atom scale on present hardware, which is large enough to
observe detailed mechanisms of yield, plastic flow and shock
interaction with nanostructures, directly.\cite{Kadau2006,Shekhar2013}
Work with empirical potentials can give important insight and
understanding, but a need for first-principles methods such as Density
Functional Theory (\DFT) exists in providing predictive power and
accuracy. These methods must use more modest system sizes, of hundreds
or thousands of atoms in the case of \DFT.

Silicon has a rich phase diagram, with metallic dense phases rather
different in character to the ambient diamond phase, making it an
interesting and challenging object of simulation.  In total, eleven
stable or metastable phases of silicon are currently
known.\cite{Mujica2003} Shock experiments have provided important data
for constructing the phase diagram.  The phase transition in silicon
from the cubic diamond structure to the beta-tin structure, occurring
at \SI{12}{GPa} at room temperature, and undergoing a reduction in
volume of 20\%, has been well established by static loading
experiments from the 1960s onward.\cite{Minomura1962,Jamieson1963}
Evidence of at least one phase transition at similar pressures was
then observed in shock-wave experiments, starting with
\citet[.][]{Pavlovskii1968}

If a shock wave is strong enough to cause a material to yield
plastically or undergo a phase transition, the wave can split into two
or more separate shock waves, and this has long been observed and
understood.\cite{Duvall1977} In this situation, the last shock takes
the material to its final state, but the preceding shocks take the
material to a cusp on the pressure-volume Hugoniot locus caused by a
transition: either the Hugoniot elastic limit or the onset pressure of
a phase transition.  In silicon, \citet{Gust1970,Gust1971} found a
three-wave structure for samples shocked in the \dirnfamily{100}
crystal direction and a four-wave structure when shocked in the
\dirnfamily{110} or \dirnfamily{111} directions.  In the latter cases,
these waves were attributed to: an initial elastic precursor to the
Hugoniot elastic limit of \SI{5.5}{GPa}, followed by waves
corresponding to a state of plastic yield and two successive phase
transitions at \SI{10}{GPa} and \SI{13}{GPa}.  Along
\dirnfamily{100}, the higher elastic limit of \SI{9}{GPa} obscures
the first transition wave, and a single wave takes the material
simultaneously to a new phase and to a state of hydrostatic stress.

The work of \citet{Goto1982} largely confirmed the findings
of \citet[,][]{Gust1971} although they observed a three-wave structure,
regardless of crystal orientation, consistent with only a single phase
transition at \SI{13}{GPa}.  Above the Hugoniot elastic limit, shock
compression was found to result in a hydrostatic stress configuration,
due to the complete loss of strength in the material.

More recently, and contrary to the earlier experimental work,
\citet{Turneaure2007,Turneaure2007x} reported a single phase
transition that is complete by \SI{15.9}{GPa}.  Shocks to these
pressures show a much greater volume compression than the points
attributed to an extended mixed-phase region by both \citet{Gust1971}
and \citet{Goto1982} Here the phase transition is not complete until
at least \SI{30}{GPa}.  This discrepancy is explained by
\citet{Turneaure2007x} as arising from the reflection of the first two
shock waves propagating back into the material before the arrival of
the third wave, and altering the peak state of the earlier
experiments.  They avoid this eventuality by backing the silicon with
a window made from lithium fluoride, a material with a good impedance
match to silicon.

The \textit{Imma} phase of silicon is found intermediate between the
beta-tin and simple hexagonal phases, and is stable between
\SI{13}{GPa} and \SI{15}{GPa} at room
temperature.\cite{McMahon1994} Theoretically, the energy and volume of
these three phases are close.\cite{Mujica2003} A recent simulation of
directly shocked silicon using an empirical potential\cite{Mogni2014}
found a phase transition to an \textit{Imma} phase with a modification
of the Tersoff potential\cite{Tersoff1986,Tersoff1988} of
\citet[.][]{Erhart2005}

In this paper, we give the Hugoniot loci according to Density
Functional Theory for several pure phases of silicon, including cubic
diamond under elastic compression along \dirnfamily{100},
\dirnfamily{110} and \dirnfamily{111}, a hydrostat (resulting from
either a single shock or a split-shock structure), beta-tin, simple
hexagonal and the liquid, and report shock temperatures for these
states.

Several approaches can been taken for the determination of a Hugoniot
locus from molecular dynamics.  The most straightforward, but
computationally the most demanding, is to simulate a slab of atoms
struck by an impactor directly, measuring the speed of any shock waves
and post-shock average particle velocities as they arise from the
simulation.  From the Hugoniot relations, these velocities can be
converted to a relationship between pressure and volume compression.
For empirical potentials, a local stress is conveniently available, so
this could also be taken directly from the simulation.  This is the
approach taken by, e.g. \citet{Kadau2007}

It is simple to check that a given equilibrium state lies on or close
to the (single-shock) Hugoniot locus, which amounts to satisfying
the Hugoniot relation
\begin{equation}
E- E_0 = -\half(\sigma^{33} + \sigma^{33}_0)(V_0 - V),
\label{eq:ehug}
\end{equation}
where $E$ is the internal energy, $V$ is the specific volume and
$\sigma^{33}$ is the stress in the direction of the shock (and can be
replaced with the pressure $p$ in a hydrostatic situation).  The
zero-subscripted variables are for the pre-shocked state.  Other
(equivalent) Hugoniot relations exist between any three of: internal
energy, pressure, volume, shock velocity and particle velocity.  It is
therefore sufficient to sample several points that are chosen to
bracket the Hugoniot locus, and the Hugoniot state then approximated
by interpolation, or solved for iteratively.  The former is the
approach taken by \citet{Galli2004} for shocked deuterium.

\citet{Swift2001} constructed a polymorphic equation of state for
silicon, incorporating \DFT simulations of the cubic diamond and
$\beta$-Sn phases, with the lattice-thermal contribution approximated
by quasiharmonic phonons.  The equation of state was constructed with
a particular focus on simulating shock waves.  The full equation of
state was sampled and the Hugoniot locus could therefore be extracted
as a one-dimensional path through it.  The phase boundary and mixed
phase region along the Hugoniot were found explicitly by minimizing
the Helmholtz free energy computed from the quasiharmonic phonon
approximation.

Alternatively, a Hugoniot state can be determined dynamically from
within a single molecular dynamics simulation by some modified
dynamics to constrain the state to satisfy \cref{eq:ehug}.  This is
the approach taken by the Hugoniostat
methods\cite{Maillet2000,Ravelo2004} and the technique of
\citet{Reed2003} The former simulations use modified Nos\'e--Hoover
dynamics while the latter uses coupled dynamics of the atoms and
simulation cell, whose Lagrangian involves the computed instantaneous
shock speed, and varies the simulation cell uniaxially.  One aim of
these dynamics is to work on timescales comparable to shock-passage
times, without the overhead of dealing with a direct non-equilibrium
simulation.

If we are interested only in the final post-shock state, and are not
interested in the (modified) dynamics while the constraint is being
applied, we are free to use a method based on simple velocity
rescaling, analogous to the procedure of
Berendsen,\cite{Berendsen1984} which is what we propose here due to
its increased efficiency in reaching the final state.

\section{Computational Method}
\label{sec:method}
\subsection{Density Functional Theory}
The ab initio \MD simulations described here were performed with the
\SIESTA method and implementation of Density Functional Theory,
\cite{Soler2002} using the \citet{Perdew1996} \textsc{gga}
functional.

\begin{table}
\begin{ruledtabular}
  \caption{Basis parameters for silicon, according to the
    soft-confinement scheme of \protect\citet{Junquera2001}.  For the
    purposes of basis generation, an effective ionic charge of
    -0.46 was used, which was also variationally optimized.  The
    cutoff radii of the first and second zeta functions are
    $r(\zeta_1)$ and $r(\zeta_2)$, and $r_i$ is the confinement
    potential's internal radius.  $V_0$ is the soft-confinement
    prefactor.}
\begin{tabular}{rrcccc}
$n$ &$l$  &$r_i$ ($a_0$) &$r(\zeta_1)$ ($a_0$) &$r(\zeta_2)$ ($a_0$) &$V_0$ (Ry)  \\
\hline
3   & 0   & 4.97         & 7.00                & 4.38               & 15.43 \\
3   & 1   & 3.83         & 7.00                & 4.09               & \phantom{0}4.70 \\
3   & 2   & 0.03         & 4.55                & -                  & 11.97 
\end{tabular}
\label{tab:basis}
\end{ruledtabular}
\end{table}

The core electrons were described with a Troullier--Martins
norm-conserving pseudopotential\cite{Troullier1991} with a matching
radius in each angular momentum channel of \SI{1.89}{a_0}.  The
valence electrons were described with a basis of numerical atomic
orbitals of double-$\zeta$ polarized type\cite{Junquera2001}
(representing 13 orbitals per atom).  The basis was generated by
fixing the longest orbital cutoffs at \SI{7.0}{a_0} and
variationally optimising the other parameters in bulk diamond-phase
silicon---the final basis parameters are given in \cref{tab:basis}.

The mesh used for integrals in real-space was well converged at a grid
cutoff of \SI{100}{Ry}.  The dense phases of silicon required
several $k$-points to converge in energy, and in particular, for the
cold compression curves of the various phases to converge in energy
relative to one another.  A $4^3$ Monkhorst--Pack grid of points was
used on the 64 atom simulations, to give an effective cutoff length of
\SI{21}{\angstrom}.

The electronic temperature used in the \DFT calculations should be
consistent with the final temperature attained after the annealing
process described below.  The consistent forces for the ab initio
molecular dynamics are the nuclear-position derivatives of the
electronic free-energy as defined in Mermin's \DFT.\cite{Mermin1965}
All of the simulations reported below are for an electronic
temperature of \SI{300}{K}, except for the two points with highest
temperatures, for which the electronic temperature was adjusted to
coincide with the final (nuclear) temperature.  The effect of the
electronic temperature on the reported quantities was found to be
quite small: the maximum difference in pressure for the hottest
simulation between using a consistent electronic temperature and the
initial \SI{300}{K} is below 5\%.

The integration of the dynamics used the Born-Oppenheimer
approximation with a timestep of \SI{1}{fs}.

\subsection{Annealing to the Hugoniot Locus}
We use a simple annealing procedure to find the state on the Hugoniot
corresponding to a specified longitudinal strain.  A Berendsen
thermostat\cite{Berendsen1984} is used with a variable target
temperature computed from the instantaneous difference in energy
between the total energy of the system, and the total energy that
would be required to satisfy the energy Hugoniot relation,
\cref{eq:ehug}, exactly, given the current instantaneous longitudinal
stress.

The procedure is given explicitly below.  This may be combined with a
further anneal to relax the pressure to a hydrostatic configuration if
desired.  Optionally, the box vectors may be gradually ramped between
two states, which is most useful when the starting state of the
simulation and the initial state of the Hugoniot locus are the same.

\begin{algorithmic}
\Procedure{HugoniotAnneal}{$E^{\text{tot}}_0, V_0, \sigma_0$}

    \State compute $E^\text{tot}, \varsigma, \bm{F}_n$ from atomic positions $\bm{x}_n$

    \ForAll{atoms} \Comment velocity Verlet
        \State $\bm{v}_{n} \gets \bm{v}_{n-1} + \dfrac{\dd{}t}{2m} ( \bm{F}_{n-1} + \bm{F}_n )$
        \State $\bm{x}_{n+1}^{\text{(unc)}} \gets \bm{x}_n+\dd{}t\, \bm{v}_n + \dfrac{\dd{}t^2}{2} \bm{F}_n /m $
    \EndFor
    
    \State $\sigma \gets \frac{1}{V} \sum_{\text{atoms}} m\, \bm{v}_n \otimes \bm{v}_n + \varsigma$      \Comment compute the stress
    \State $E^{\text{kin}} \gets \sum_{\text{atoms}} \half m\, \bm{v}_n \cdot \bm{v}_n $
\\ \Comment compute the target energy
    \State $E^{\text{hug}} \gets E^{\text{tot}}_0 - \half(\sigma^{33} + \sigma^{33}_0)(V_0 - V)$     
    \State $E^{\text{kin}}_{\text{target}} \gets E^{\text{kin}} + E^{\text{hug}} - E^{\text{tot}}$
    \State $r^2 \gets \left( 1 + \dfrac{\dd{}t}{\tau_\text{relax}} \left( \dfrac{E^\text{kin}_\text{target}}{E^{\text{kin}}} - 1 \right) \right)$
 
     \ForAll{atoms}
         \State $\bm{v}_n^{\text{(sca)}} \gets r \bm{v}_n$      \Comment scale the velocities
 \\   \Comment correct positions based on the scaled velocities
         \State $\bm{x}_{n+1} \gets \bm{x}_{n+1}^{\text{(unc)}} + \dd{}t\,(\bm{v}_n^{\text{(sca)}} - \bm{v}_n)$ 
     \EndFor
    \State $t \gets t + \dd{}t$, $n \gets n + 1$
\EndProcedure
\end{algorithmic}
The meaning of the variables used is as follows. $E$ denotes an energy
(refer to the sub and superscripts), $V$ is the unit cell volume,
$\sigma$ is the stress, which is the sum of a kinetic term and the
strain derivative of the total electronic energy $\varsigma$; $\bm{x}_n$,
$\bm{v}_n$ and $\bm{F}_n$ are the atomic positions, velocities and forces
at the $n$th timestep (`unc' stands for `uncorrected' and `sca' for
scaled), $m$ is the mass of a given atom, $\tau_\text{relax}$ is the
relaxation time, $t$ and $\dd{}t$ are the current time and timestep,
and anything with a subscript `$0$' refers to its (time averaged)
value in the unshocked state (which may be different from the starting
state of the simulation).

Even though Berendsen thermo- and barostats do not reproduce canonical
statistics,\cite{Harvey1998} it is well known that they are much more
efficient at annealing to a given equilibrium state at a desired
temperature or pressure compared with modified dynamics, such as
Nos\'e--Hoover.  The same applies here, compared to the related
Hugoniostat for shocks, and this justifies their use, since we are
interested only in the outcome of the anneal, not the intermediate
dynamics.  After the time-averaged state of the system closely
satisfies the Hugoniot relation, the simulation can be restarted with
Verlet dynamics to check if \cref{eq:ehug} is indeed satisfied.

\begin{figure}
\includegraphics{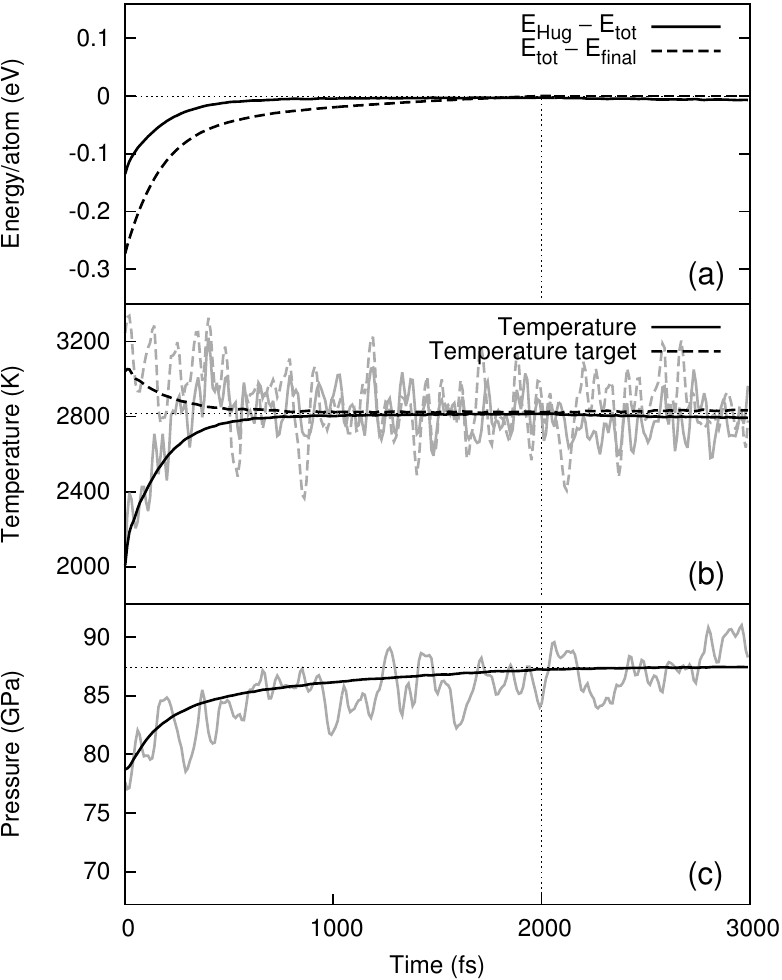}
\caption{Response of (a) internal energy and the difference with
  the Hugoniot energy computed from \cref{eq:ehug}, (b) temperature
  and target temperature, and (c) pressure, to the Hugoniot anneal
  described here with a relaxation time of \SI{100}{fs}.  After
  \SI{2000}{fs}, the anneal is switched off and the dynamics
  continued with Verlet integration.  The response is averaged over
  \num{10000} independent 216 atom Stillinger--Weber silicon systems,
  starting from a \SI{2000}{K} liquid and annealed to the Hugoniot
  locus with an initial state of \SI{300}{K} and zero stress.  For
  comparison, the light grey lines are taken from a single
  trajectory---in the energy plot, this is indistinguishable from the
  mean.}
\label{fig:ehugdiff}
\end{figure}

\Cref{fig:ehugdiff} shows the convergence in total energy, temperature
and pressure of liquid Stillinger-Weber silicon to a state on the
particular Hugoniot locus from an initial state of \SI{300}{K} and
zero pressure.  This is an averaged result of \num{10000} independent
simulations, each starting in the liquid phase at \SI{2000}{K}.  The
relaxation time used was \SI{100}{fs}.  After \num{2000} timesteps of
\SI{1}{fs}, the anneal is switched off and Verlet integration used for
the remaining time.  Note the slight relaxation of temperature and
pressure away from their final values under the thermostat.  For this
case, it amounts to a temperature difference of within~1\%.

\section{Results}
\label{sec:results}
\begin{figure}
\centering 
\includegraphics{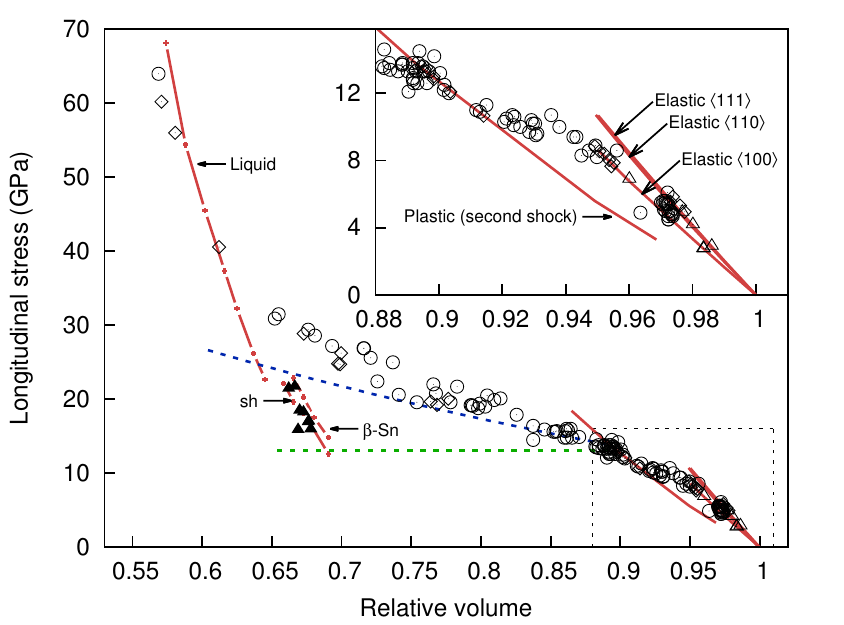}
\caption{Pressure--volume Hugoniot loci for silicon.  The solid red
  lines in the figure are the \DFT results from this work (with
  contained points indicating individual simulations), with an initial
  pre-shocked state of zero pressure and \SI{300}{K}, with the final
  state in the indicated phase (`sh' for simple hexagonal).  Estimated
  error is less than 5\% for the liquid and beta-tin phases, and is
  substantially smaller for the diamond phase.  The symbols are
  experimental results from the literature: $\circ$
  \protect\citet{Gust1971}, $\diamond$ \protect\citet{Goto1982},
  {\tiny{$\triangle$}} \protect\citet{Turneaure2007}, {{{\scriptsize
        {$\blacktriangle$}}}} \citet{Turneaure2007x}.  The dashed
  lines are approximations to the mixed-phase portion of the Hugoniot,
  for cubic diamond to: liquid (upper blue line) and beta tin (lower
  green line).}
\label{fig:hug-siesta-expt-si-pv}
\end{figure}

\begin{figure}
\centering 
\includegraphics{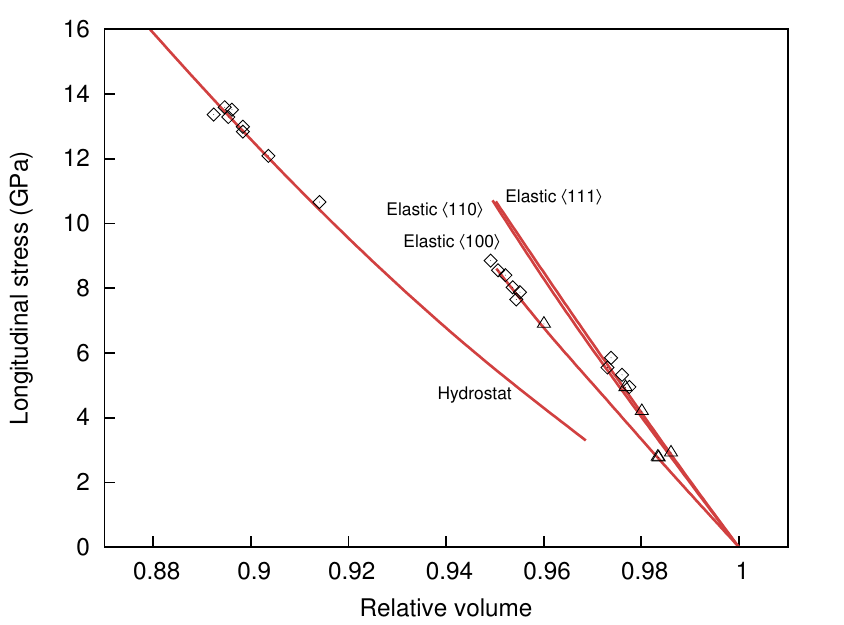}
\caption{Pressure--volume Hugoniot loci for silicon.  This is a
  similar plot to \cref{fig:hug-siesta-expt-si-pv}, with the meaning
  of the symbols and lines the same, emphasizing the small strain
  region of the Hugoniot locus and with the results of
  \protect\citet{Gust1971} omitted due to their larger variance.}
\label{fig:hug-siesta-expt-si-pv-inset}
\end{figure}

\begin{figure}
\centering
\includegraphics{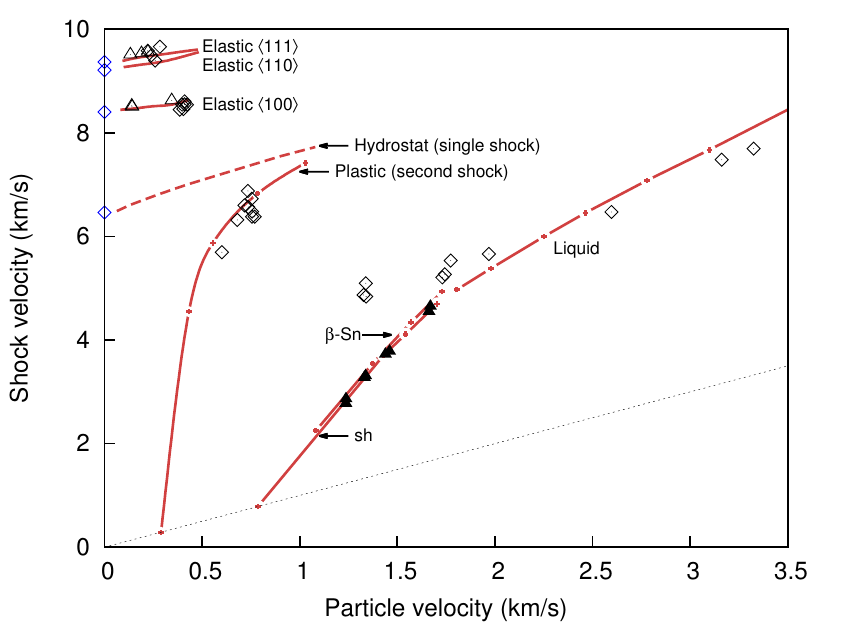}
\caption{Particle velocity--shock velocity Hugoniot loci for silicon.
  The \DFT results (solid red lines and points) each correspond to an
  initial state of zero pressure and \SI{300}{K}, with the final state
  in the indicated phase.  The dashed line is for a single-shock
  process whose final state has a hydrostatic stress configuration.
  The meaning of the symbols is the same as in
  \cref{fig:hug-siesta-expt-si-pv}, with the blue diamonds on the axis
  the elastic and bulk wave speeds from \citet{Goto1982} The dotted
  base line indicates equal shock and particle velocity, below which
  no viable shock should be recorded.}
\label{fig:hug-siesta-expt-si-us-up}
\end{figure}

\begin{figure}
\centering
\includegraphics[width=0.5\textwidth]{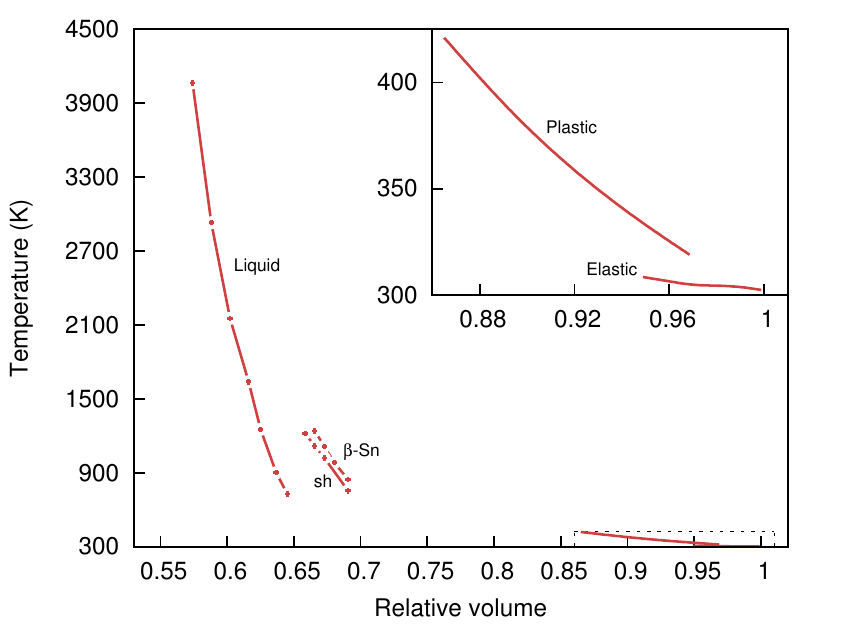}
\caption{Post-shock temperature as a function of volume for several
  final states.  The \DFT results (solid red lines) each correspond to
  an initial state of zero pressure and \SI{300}{K}, with the final
  state in the indicated phase.  The `plastic' curve does not include
  the temperature rise due to dissipative heating.  The meaning of the
  symbols is the same as in \cref{fig:hug-siesta-expt-si-pv}}
\label{fig:shock-temperature}
\end{figure}

The calculated pressure--volume and shock-velocity--particle-velocity
Hugoniot loci for the pure phases are compared to results from several
experiments in
\cref{fig:hug-siesta-expt-si-pv,fig:hug-siesta-expt-si-pv-inset,fig:hug-siesta-expt-si-us-up}.
The specific volume at zero pressure and \SI{300}{K} for the
\textsc{pbe} functional is \SI{0.421}{cm^3/g}, which is smaller than
the experimental value of \SI{0.431}{cm^3/g}.  The reduced volume is
plotted in the figures: if the specific volume were plotted instead,
the \textsc{dft} results would be offset by an amount corresponding to
the difference in zero-pressure volume.  The particle velocity--shock
velocity Hugoniot is not greatly affected.

The curves for the elastic shocks are computed from a uniaxial box
deformation along the indicated direction.  The `plastic' curve is for
a split shock, with an elastic precursor to \SI{6}{GPa}, taking the
material to a hydrostatic stress configuration: this supposes that the
material has no residual strength.  The hydrostat in
\cref{fig:hug-siesta-expt-si-us-up} is for an unphysical shock process
that relaxes the material to hydrostatic stress  behind
a single, unsplit shock wave.  This permits comparison with the bulk
speed of sound (the shock velocity for this wave should extrapolate to
the bulk speed of sound at zero particle velocity.)

When comparing the hydrostat and the `plastic' curve to the yielded
phase, we assume that the yielding serves only to remove the
deviatoric stress, and that the bulk response of the material is
unaffected.  We neglect the dissipative heating due to this effect.

The agreement with the experimental data for the elastic and plastic
shocks is good, with the compressibility along \dirnfamily{100},
\dirnfamily{110} matching well in value and \dirnfamily{111} showing
the correct trend (although underestimating the value).  The close match
between the experimental plastic shock pressures and the hydrostatic
plastic shock calculated here supports the observation that the
material loses all of its strength after yield.

\begin{table}
\begin{ruledtabular}
  \caption{Coefficients of a linear fit the shock velocity for the
    elastic waves, $U_s = c_0 + su_p$, for this work and two sets of
    experimentally-determined values.}
\begin{tabular}{lcccccccc}
          & \multicolumn{2}{c}{\dirnfamily{100}} & \multicolumn{2}{c}{\dirnfamily{110}} & \multicolumn{2}{c}{\dirnfamily{111}} & \multicolumn{2}{c}{Bulk} \\
          & $c_0$ & $s$ & $c_0$ & $s$ & $c_0$ & $s$ & $c_0$ & $s$ \\
          & (\si{km/s}) & (-) & & & & & &\\
\hline
This work & 8.38 & 0.42 & 9.21 & 0.57 & 9.34 & 0.57 & 6.51 & 1.18 \\
Ref.\ \onlinecite{Goto1982} & 8.42 & 0.32 & 9.24 & 1.01 & 9.39 & 0.98 & \multicolumn{2}{c}{-} \\
Ref.\ \onlinecite{Hall1967}&8.43 & - & 9.13 & - & 9.34\footnotemark[1]& - & 6.48\footnotemark[1] & -\\
\end{tabular}
\label{tab:lin-coeffs}
\footnotetext[1]{Calculated from the given elastic constants and density.}
\end{ruledtabular}
\end{table}

The particle and shock velocities in
\cref{fig:hug-siesta-expt-si-us-up} are computed from the computed
pressure and volume points using the Hugoniot relations
\begin{align}
u_p^2 &= (p-p_0)(v_0-v)\\
U_s^2 &= v_0^2 (p-p_0)/(v_0-v),
\end{align}
where $v_0$ and $v$ are the initial and final specific volumes.  A
linear fit to the elastic part of the
shock-velocity--particle-velocity Hugoniot has coefficients given in
\cref{tab:lin-coeffs}.  The extrapolated value of the bulk sound speed
of \SI{6.51}{\kilo\meter\per\second} agrees very well with the value
of \SI{6.48}{\kilo\meter\per\second} calculated from the second
order elastic constants.\cite{Gust1971,Hall1967}

The $\beta$-Sn and simple hexagonal curves each correspond to a three
wave split shock structure, behind an elastic wave to the experimental
elastic limit of \SI{6}{GPa} and a secondary wave to the
experimental location of the phase transition at \SI{13.8}{GPa}.
For both of these waves, the computed volume for the \dirnfamily{100}
direction was used for the post-shock state.  In general, it is quite
insensitive to the precise location of the wave split, particularly
for the elastic case, since the contribution to the energy change is
much smaller than the 20\% volume reduction across the phase change.
The final stress was hydrostatic.  Since the $c/a$-ratio is free in
the $\beta$-Sn and simple hexagonal structures, an additional
relaxation step was used on the simulation box to impose a hydrostatic
distribution of stress while simultaneously annealing to the Hugoniot.
The $\beta$-Sn and simple hexagonal curves are close in pressure,
temperature and shock velocity, with the experimental values closest
to the simple hexagonal \DFT Hugoniot.  The computed pressures and
temperatures of these points put them in stable region for the simple
hexagonal structure on the silicon phase diagram.\cite{Kubo2008}

Part of the liquid Hugoniot corresponds to a three-wave shock
structure, with the third wave reaching the final liquid state, behind
a secondary wave to the onset of the melting transition and an elastic
precursor wave.  For the highest pressures, where the final wave has a
velocity greater than that of the secondary wave of
\SI{6.83}{km/s}, it instead corresponds to a
two-wave structure (behind only the elastic precursor).  The largest
shock pressures closely match the calculated liquid Hugoniot, with the
simulated liquid being systematically slightly too stiff.

The predicted post-shock temperatures (given in
\cref{fig:shock-temperature}) indicate that these highest pressure
points are likely to be liquid phase.  The sixfold coordinated
liquid lies close in $p$--$v$ to the Hugoniot for the beta-tin phase,
and so this phase transition does not exhibit the large mixed phase
region as for the diamond to dense-phase silicon.

\subsection{The Phase Transition}
There is a considerable range of relative volume between the Hugoniot
loci of the pure phases shown in \cref{fig:hug-siesta-expt-si-pv}.
The experimentally measured points in this region have a final state
that is a mixture of two phases.  Points on the mixed-phase region of
the Hugoniot are on the intersection of the phase boundary for the two
phases, as well as satisfying \cref{eq:ehug}.

Similar to the plastic shock, a pressure--volume Hugoniot is convex at
the onset of a mixed phase region: if the change in slope is great
enough, this causes the shock to split into a wave taking the material
to the pressure at the onset of the phase transition, and a slower
wave taking the material to its final state, which is a coexistence of
the two phases.

The Hugoniot locus through the mixed phase region can be constructed
by considering the jump condition in enthalpy across the shock from
the point (`1') at the onset of the transition to a point (`2') on the
mixed Hugoniot
\begin{equation}
  H_2 - H_1 = E_2 - E_1 + p_2V_2 - p_1V_1,
\end{equation}
and on substituting \cref{eq:ehug} for the jump in internal energy,
this reduces to
\begin{equation}
  H_2 - H_1 = \half (p_2 - p_1)(V_2 + V_1).
\end{equation}
The latent heat $L$ of the phase transition results in a change in
enthalpy, written according to the Clausius--Clapeyron equation as
\begin{equation}
\lambda L = -T \deriv{p}{T} (V_1 - V_2),
\end{equation}
where $\lambda$ is the mass fraction of the second phase and the
derivative is along the phase line.  

Since the mixed region is not at constant pressure, there is an
additional contribution to the enthalpy change from the difference in
pressure and volume between the onset of the transition and the
post-shock state.  This leads to a linearized equation relating the
pressure and volume changes on the phase-transition
shock,\cite{Duff1957}
\begin{multline}
p_2 - p_1 = (V_1 - V_2)\times \\\left[\beta V_1 + \left(\frac{1}{2T_1}(V_1 - V_2)-2\alpha 
    V_1\right)\deriv{T}{p}  + \frac{C_p}{T_1}
  \left(\deriv{T}{p}\right)^2\right]^{-1},
\end{multline}
where $\beta$ is the isothermal compressibility, $\alpha$ is the
volumetric thermal expansion coefficient and $C_p$ is the heat
capacity at constant pressure.  The derivative $\dd{T}/\dd{p}$ is once
again along the phase boundary.

We require knowledge of the onset of the transition in the $p$--$V$
plane, which is not available from the single phase simulations alone
(the simulated materials are capable of being substantially superheated or
supercooled).  This could be obtained from the point
where the Hugoniot cuts the phase boundary obtained by some other
method.  

\begin{table}
\begin{ruledtabular}
  \caption{Summary of values used at the onset of the cubic diamond to
    liquid phase transition.  The phase line is as obtained by the
    experiment of \citet{Kubo2008}.  The other values are from \citet{Hull1999},
    with $\alpha$ and $c_p$ at \SI{1600}{K} and ambient pressure, and
    $\beta$ at \SI{298}{K} and \SI{13.8}{GPa}.}
\begin{tabular}{ccccc}
 $T$ (\si{K}) & $dT/dP$ (\si{K GPa^{-1}}) & $\alpha$ (\si{K^{-1}}) & $\beta$ (\si{GPa^{-1}}) & $c_p$ (\si{Jg^{-1}K^{-1}})\\
\hline
1683 & 62.4 &\num{4.5E-6} & 0.024 & 1.0 
\label{tab:phase-constants}
\end{tabular}
\end{ruledtabular}
\end{table}

We consider here two possible phase transitions starting from silicon
in the cubic diamond structure: to a liquid, and to the beta-tin
structure.  In addition, we assume that the onset of either transition
occurs at \SI{13.8}{GPa}, close to the observed experimental value.
The phase lines are experimental values, obtained by \citet{Kubo2008}
This gives the two dashed lines appearing in
\cref{fig:hug-siesta-expt-si-pv}.  The lower, green dashed line for
diamond structure to beta-tin is nearly at constant pressure, since
its slope is dominated by the steep phase-line of the
transition,\cite{Kubo2008} $dT/dp = \SI{-1426}{K/GPa}$.  This is
consistent with the experiment of \citet[.][]{Turneaure2007x} The
upper, blue dashed line for melting the diamond structure is
influenced most strongly by the compressibility $\beta$ of the cubic
diamond phase at the pressure and temperature of the onset.
Representative literature values for the constants appearing in the
above expression for the liquid are summarized in
\cref{tab:phase-constants}.  This line underestimates the
experimentally observed slope seen by \citet{Gust1971} and
\citet{Goto1982} While the simulated temperature at this pressure is
much too low for melting, the simulations of the `plastically-yielded'
state do not include dissipative heating and this could cause a
considerable temperature rise above those reported in
\cref{fig:shock-temperature}.

\vspace{0.5cm}

\section{Conclusion}
\label{sec:conclusion}
In conclusion, we have described a simple annealing method and shown
that it may be used to obtain a state on the Hugoniot locus of a pure
phase of a material with several condensed phases efficiently, from
first-principles.  An approximation relying on the slope of the
phase boundary can be used to obtain the part of the Hugoniot
corresponding to coexistence between two phases.

In the case of silicon, the results computed using this procedure
with the forces described using density functional theory match
existing experimental data very well for pressures up to
\SI{60}{GPa}, the limit of available experimental data.  We have
provided a prediction of the shock temperatures of silicon over this
pressure range.  This study supports the conclusions of the
experimental work in general, that silicon after yield supports no
deviatoric stress, and of \citet[,][]{Turneaure2007x} that the first
observed phase transition along the shock locus is likely to be to
simple hexagonal. 

\begin{acknowledgements}
This research was supported with funding from Orica Ltd.\ and the
following grants: MINECO-Spain's Plan Nacional Grant No.\
FIS2012-37549-C05-01, Basque Government Grant PI2014-105 CIC07
2014-2016, EU Grant ``ElectronStopping'' in the Marie Curie CIG
Program.  Part of this work was performed using the Darwin
Supercomputer of the University of Cambridge High Performance
Computing Service (\url{http://www.hpc.cam.ac.uk/}), provided by Dell
Inc.\ using Strategic Research Infrastructure Funding from the Higher
Education Funding Council for England and funding from the Science and
Technology Facilities Council. We thank Alan Minchinton, Richard
Needs, Nikos Nikiforakis, Stephen Walley and David Williamson for
useful input and discussions.
\end{acknowledgements}

\end{document}